# TO THE IONIZATION COOLING IN A RF CAVITY WITH ABSORBER

Alexander A.Mikhailichenko, Cornell University, LEPP, Ithaca, New York

*Abstract*

We are considering a RF cavity with Beryllium disk installed in the middle of the cavity as an ionization cooling element for the muon/pion beam. Specially arranged wedge-type shape of the disk together with nonzero dispersion allows 6D cooling of muon beam. Technical aspects of this system and conceptual design are discussed in this paper also. This type of cooler demonstrates advantages if compared with the RF cavity filled with pressurized gas or with the helical cooler.

## OVERVIEW

Ionization cooling of particles proposed a long time ago in [1]. Basic concept of this proposal is that the friction force acting to the particle moving through the absorber, directed against instant velocity of particle. As the longitudinal component of momentum lost in absorber could be restored by the longitudinal electric field in a RF cavity, the loss of transverse component is not, so this process resulting emittance reduction. This process is similar to the one with radiation losses; in some sense excitation of betatron oscillation while emitting the quanta in a channel with nonzero disprsion is similar to the (multiple) scattering in absorber.

Further this concept was developed in [2], [3]. In the preceding publication multiple scattering of cooled particles was considered together with radiation.

Later on, the ionization cooling method was found to be useful for cooling of *muons* rather than protons [4], [5].

Different combinations of Ionization loss media and RF cavities sequences considered in a framework of technical realization of this idea [6]-[14]. Main problem with ionization cooling is the scattering of cooled particles in the material of absorber.

It can be shown (see for example [14]) that the rate of change of normalized emittance $\varepsilon_\perp$ and the energy spread $\Delta E = E - E_0$ around mid-energy $E_0$ are

$$\frac{d\varepsilon_\perp}{ds} \cong -\frac{\varepsilon_\perp}{\beta^2 E}\frac{dE}{ds} + \frac{\beta_\perp \cdot (13.6 MeV/c)^2}{2\beta^3 E m_\mu X_0}, \quad (1)$$

$$\frac{d(\Delta E)^2}{ds} \cong 2(\Delta E)^2 \frac{d}{dE}\left(\frac{dE_0}{ds}\right) + \frac{d(\Delta E)^2_{str}}{ds}, \quad (2)$$

where $\beta_\perp$ stands for the betatron function, $E$ is the particle energy, $X_0$ is radiation length, the last term in (2) describes the straggling. At the equilibrium $d\varepsilon_\perp/ds = 0$, so the equilibrium emittance, according to (1) could be reached is

$$\varepsilon_{\perp equ} = \frac{\beta_\perp \cdot (13.6 MeV/c)^2}{2\beta m_\mu X_0 \cdot (dE/ds)} \quad (3)$$

The idea which we suggest in this current publication is close to the one which appointed the pressurized cavity with Hydrogen gas [15]. We are suggesting reduction of negative effect of the scattering by installation of an absorber inside a RF cavity at a midplane, while external focusing system makes the absorber location at a crossover of envelop function.

## THE CONCEPT

The absorber installed in the middle of the cavity; in the minimum of the envelop function. The last arranged with the help of pair of short focusing lenses, so the absorber located between the lenses. In a simplest case the solenoidal lenses can be used here. Focal distance of such lens is

$$F \cong \frac{4(HR)^2}{\int H^2(s)ds}, \quad (4)$$

where $(HR)$ is the magnetic rigidity of particle, $H(s)$ is the axial field value along the longitudinal coordinate $s$. It is obvious that the focal distance should equal to a half of the distance between lenses, $L=2F$ if the only pair of lenses is present. In case of periodic system the focal distance should satisfy th condition $L=4F$, i.e. the focusing should be twice of previous. The value of envelope function $\beta_\perp$ could be estimated by $\beta_\perp \cong 2cP_s/eB_s$, where $P_s$ is longitudinal component of momenta.

Location of any conducting material inside the cavity is not a comment practice, however. Main point of concern is perturbation of fields and heating. From the other hand metallic walls of the cavity, exposed to the electric and magnetic RF field is well accepted concept.

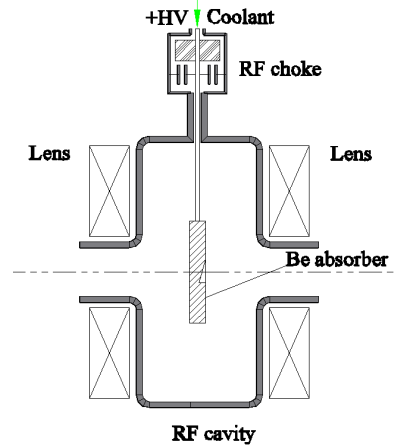

Figure 1. The concept of cooling device. The lenses produce a crossover in the mid-plane of RF cavity. The wedge-like grove is seen at the center of absorber.

Meanwhile perturbation of RF fields and losses can be made small. First example gives the ordinary pillbox cavity operating with $E_{010}$ mode, which frequency does not depend of the height of the cavity. This means that the thin disk can be installed across the cavity without perturbation of fields (and frequency). Also, it is well known practice of installation of anti-multipactoring ring-type electrodes inside the cavity with application of HV DC potential.

Basic concept of the cooling unit represented in Fig.1. Beryllium disk (absorber) is supported by the radial spokes (see Fig.1), with the coolant running inside. These spokes are insulated from the walls of cavity and HV voltage applied to the disk for prevention of multipactor (this is not important for the method proposed, however). For reduction of resistive losses the central disk is galvanized by Copper for better conductance. To prevent the leakage of RF the outer end of the supporting tubes are equipped by capacitors and inductances (choke filters). The absorber plate can any shape, we considering a thin cylindrical one. At the center the disk has an edge groove. This is made for the possibility of 6D cooling, if the dispersion separation of particles with different energies generated by optical channel. As in absence of bending magnets the dispersion propagates in optical system in the same manner (same matrices) as the envelope function this condition could be supported at all channel.

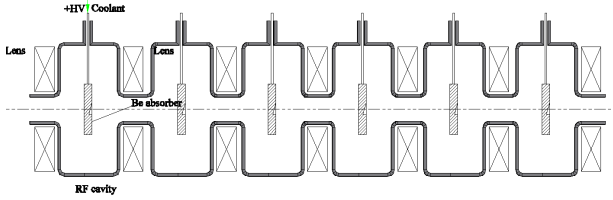

Figure 2: The cooling system assembled as an accelerating structure. Cavities can be excited individually as well as by the coupling through the central drift tubes.

We envision the normal conducting cavities so they can operate in strong magnetic field. The frequency of this structure is long 250MH-1 GHz. We suggest either π/2 mode for operation of structure or even independent excitation of cavities. Influence of multiple scattering should be compared with the angular spread in the beam generated in a target

$$\frac{21}{pc}\sqrt{\frac{\delta}{lX_0}} \sim \sqrt{\frac{\gamma\varepsilon}{\beta\gamma}} \qquad (5)$$

It is easy to compare the effectiveness of Be absorber vs pressurized $H_2$ as the radiation length of Be~35.3 cm and for the Hydrogen pressurized for 100 *atm* it comes to 8000cm, so the length of 40 cm of pressurized Hydrogen is equivalent of 1.75 mm Be absorber only.

## MODEL

We investigated possibilities for such cavity here with numerical modeling. As this is a 3D system, we used appropriate codes able to carry 3D RF calculations. We made a model for calculation with 3D code FlexPDE [16]. Absorber in this model supported with the help of two spokes.

Excitation of cavity is going through the top and bottom surfaces, where the electric field strength was taken as

$$E_z(\rho, z_{in/out}) = E_0 \cdot J_0(2.405 \cdot \rho/R_{cav}) Cos(\omega t) \qquad (6)$$

The frequency in this model was varied until the maximum was found (with some accuracy).

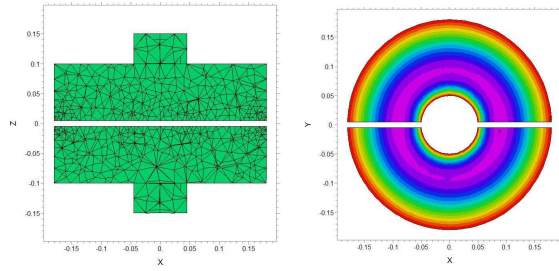

Figure 3. The mesh (at the left) ; Electric field modulus in a median plane cut.

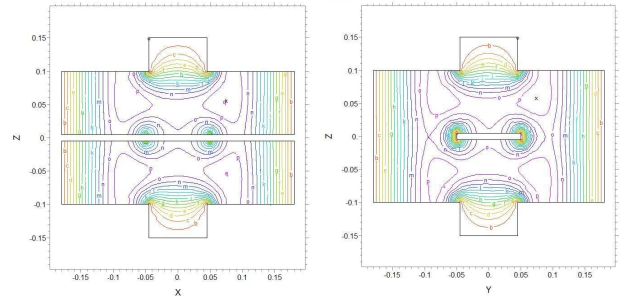

Figure 4. The lines of constant field strength in two orthogonal projections.

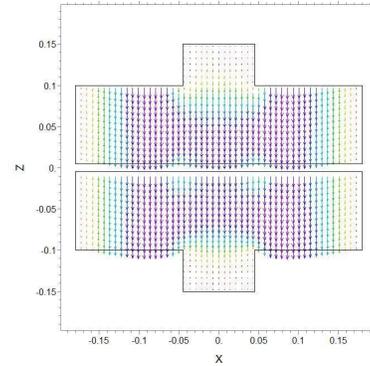

Figure 5. Electric vector field (at the left);

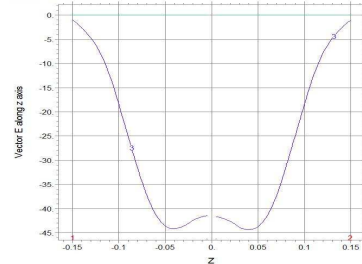

Figure 6: Electric field as a function of longitudinal coordinate.

## DESIGN CONCEPT

Design concept of one cell is represented in Fig. 7. The central disc supported in this design by three spokes made with hollow tubes carrying the coolant. A wedge type groove is seen here at central region.

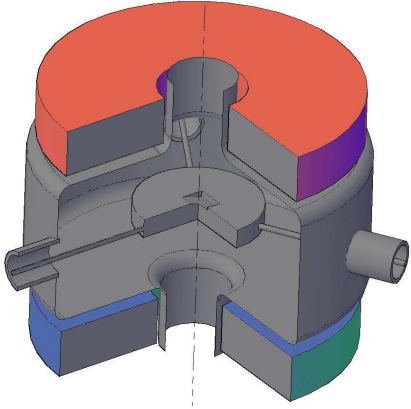

Figure 7: 3D sketch of the system represented in Fig.1.

Electrical insulation of absorber and application of HV for prevention of multipactor is optional.

## FOCUSING COILS

We suggest alternating fields in neighboring lenses. The reason is in enlarged acceptance of such configuration and in preventing of this configuration in capturing secondary electrons (with the same polarity of currents in the coils the resulting field looks like a field of solenoid with periodic modulation.

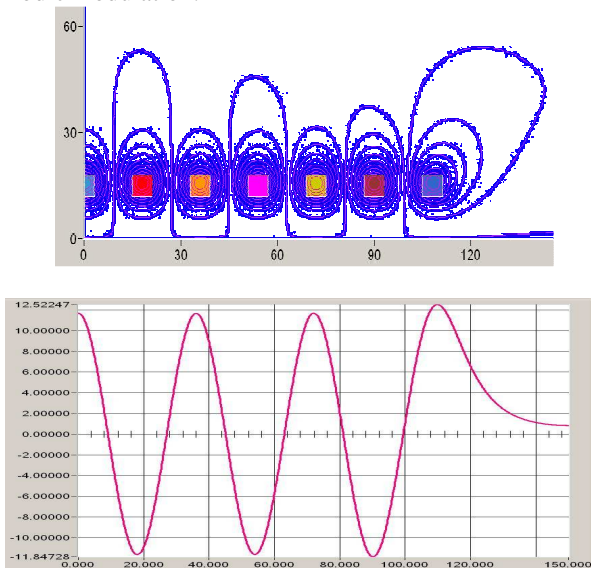

Figure 10. Magnetic field lines and the field dependence in a sequence of solenoidal lenses.

The focusing coils could be wound with SC cable. In this case the coil with moderate cross section 5x10cm$^2$ can carry total current $NI$~5 MA x turns, which can deliver the axial peak field ~ $B_s \cong 0.4\pi NI/L \sim 10T$, where the effective length of magnetic line $L$~40cm was substituted for estimation. For the particles with $pc$ ~0.3$GeV$, the magnetic rigidity comes to ($HR$)~1T x m, which delivers the focal distance (4)

$$F \cong \frac{4(HR)^2}{\int H^2(s)ds} \cong \frac{4 \cdot 1}{100 \cdot 0.5} \cong 0.08m \equiv 8cm, \qquad (7)$$

where the characteristic distance ~ 0.5 $m$ was substituted. These numbers are the optimistic ones and give the scale.

## SUMMARY

The type of cooler allows natural engineering realization of concept. The idea of this cooler is simple and straight forwarded. As the losses occur in a longitudinally compacted absorber, it is easy to arrange the waist of envelop function in the middle of the cavity, where the absorber is located, reducing influence of scattering.